\documentstyle[12pt]{article}

\begin{document}

\title{SUPERSYMMETRIC METHOD FOR CONSTRUCTING
QUASI-EXACTLY SOLVABLE POTENTIALS}

  \author{V. M. Tkachuk\\ 
  Ivan Franko Lviv State University, Chair of Theoretical Physics \\ 
		 12 Drahomanov Str., Lviv UA--290005, Ukraine\\
		 E-mail: tkachuk@ktf.franko.lviv.ua }

\maketitle

\begin{abstract}
We propose a new method
for constructing the quasi-exactly solvable (QES) potentials 
with two known eigenstates
using supersymmetric quantum mechanics.
General expression for QES potentials with explicitly known
energy levels and wave functions of ground state
and excited state are obtained.
Examples of new QES potentials
are considered.

Keywords: supersymmetry, quantum mechanics,
quasi-exactly solvable potentials.

PACS numbers: 03.65.-w; 11.30.Pb.
\end{abstract}

\section{Introduction}
A potential is said to be quasi-exactly solvable (QES) 
if a finite number of eigenstates of corresponding Schr\"{o}dinger
operator can be found exactly in explicit form.
The first examples of QES potentials were given in [1--4].
Subsequently several methods for generating QES potentials were worked out
and as a result many QES potentials were found [5--13] 
(see also review book [14]). Three different methods based respectively
on a polynomial ansatz for wave functions, point canonical
transformation, supersymmetric (SUSY) quantum mechanics are 
described in [12].

Recently, an anti-isospectral transformation
called also as duality transformation was introduced in [15] . 
This transformation
relates the energy levels and wave functions of two QES potentials.
In [16] a new QES potential was discovered using this 
anti-isospectral transformation .

The SUSY method for construction
of QES potentials was used in [10--12]. The starting point of this
method is some initial QES potential with
$n+1$ known eigenstates. Then applying the technique of
SUSY quantum mechanics (for review of SUSY quantum mechanics see
[17, 18]) one can calculate the supersymmetric
partner of the QES potential. From the properties of the
unbroken SUSY it follows that the supersymmetric partner
is a new QES potential with $n$ known eigenstates.

In addition SUSY was used to develop 
some generalized method
for the construction of the so-called conditionally exactly solvable 
(CES) potentials in [19, 20].
The CES potential is the one for which the eigenvalues problem
for the corresponding Hamiltonian is exactly solvable only when 
the potential parameters obey certain conditions.
Such a class of potentials was first considered in [21].

In the present paper 
we develop a new SUSY method for generating
QES potentials which 
in contrast to the
previous papers [10--12] does not require 
the knowledge of
the initial QES potential for the generation of a new QES one. 

\section{Supersymmetric quantum mechanics}

In the Witten's model of supersymmetric quantum mechanics
the SUSY partner Hamiltonians $H_\pm$ read
\begin{equation} \label{5}
H_\pm=B^\mp B^\pm=-{1\over2}{d^2\over dx^2}+ V_\pm(x),
\end{equation}
where
\begin{eqnarray} \label{2}
B^\pm={1\over\sqrt{2}}\left(\mp{d\over dx}+ W(x)\right), \\ \label{221}	
V_\pm (x)={1\over 2}\left(W^2(x) \pm W'(x)\right), \ \ W'(x)={dW(x)\over
dx},
\end{eqnarray}
$W(x)$ is referred to as a superpotential.
In this paper we shall consider the  systems 
on the full real line 
$-\infty<x<\infty$.

Consider the equation for the energy spectrum
\begin{equation} \label{7}
H_\pm \psi_n^\pm (x)=E_n^\pm \psi_n^\pm (x), \ \
n=0, 1, 2,... .
\end{equation}
As a consequence of SUSY 
the Hamiltonians $H_+$ and $H_-$ 
have the same energy spectrum
except for the zero energy ground state. 
The latter exists in the case of 
the unbroken SUSY. 
Only one of the Hamiltonians
$H_\pm$ has a square integrable eigenfunction corresponding to
zero energy. We shall use the convention
that the zero energy eigenstate belongs to $H_-$.
Due to the
factorization of the Hamiltonians $H_\pm$ (see (\ref{5}))
the ground state for $H_-$ satisfies the equation
$
B^-\psi_0^-(x)=0
$
the solution of which is
\begin{equation} \label{9}
\psi_0^-(x)=C^-_0\ \exp\left(-\int W(x) dx\right),
\end{equation}  
$C^-_0$ is the normalization constant. Here and below $C$ denotes
the normalization constant of the corresponding wave function.
From the condition of square integrability of wave function $\psi^-_0(x)$ 
it follows that superpotential must satisfy the condition
\begin{equation} \label{10}
{\rm sign}(W(\pm\infty)) = \pm 1.
\end{equation}
Note that this is the condition of the existence of unbroken SUSY.

The eigenvalues and eigenfunctions of the Hamiltonians $H_+$ and $H_-$
are related by SUSY transformations
\begin{eqnarray} 
&& E_{n+1}^-=E_n^+,\ \ E_0^-=0,  \label{11} \\
&& \psi_{n+1}^-(x)={1\over \sqrt{E_n^+}}B^+\psi_n^+(x), \label{12} \ \
\psi_n^+(x)={1\over \sqrt{E_{n+1}^-}}B^-\psi_{n+1}^-(x). 
\end{eqnarray}

For a detailed description of SUSY quantum mechanics and
its application for the exact calculation of eigenstates of Hamiltonians
see reviews \cite{17,18}.
The properties of the unbroken SUSY quantum mechanics 
reflected in SUSY transformation (\ref{11}), (\ref{12})
are used for exact calculation of the energy 
spectrum and wave functions.
In the present paper we use these properties for the generation of the 
QES potentials with the two known eigenstates. 

\section{QES potentials with the two known eigenstates} 

Suppose we study a Hamiltonian $H_-$, whose
ground state is given by (\ref{9}).
Let us consider the SUSY partner of $H_-$, i.e. the Hamiltonian $H_+$. 
If we calculate the ground state of $H_+$ we immediately find the first 
excited state of $H_-$ using the SUSY 
transformation (\ref{11}), (\ref{12}). 
In order to calculate 
the ground state of $H_+$ let us rewrite it in the following form
\begin{equation} \label{14}
H_+=H_-^{(1)} + \epsilon = B_1^+B_1^- + \epsilon, \ \  \epsilon > 0,
\end {equation}
which leads to the following relation between potential energies
\begin{equation}\label{171}
V_+(x)= V^{(1)}_-(x)+\epsilon,
\end{equation}
where 
$ \epsilon$ is the energy of the ground state of $H_+$ 
since we suppose that $H_-^{(1)}$ has zero energy ground state,
$B_1^{\pm}$ and $V^{(1)}_-(x)$ are given by (\ref{2}) and (\ref{221})
with the superpotential $W_1(x)$.  

As we see from (\ref{14}) the ground state wave function
of $H_+$ is also the ground state wave function of $H_-^{(1)}$ 
and it satisfies the equation
$
B^-_1\psi_0^+(x)=0.
$
The solution of this equation is
\begin{equation} \label{17}
\psi_0^+(x)=C^+_0 \ \exp\left(-\int W_1(x) dx\right),
\end{equation}  
where for square integrability of this function
the superpotential $W_1(x)$ satisfies 
the same condition as $W(x)$ (\ref{10}). 
Using (\ref{11}) and (\ref{12}) we obtain the energy level
$E_1^-=\epsilon$
and the wave function of the first excited state 
$\psi_1^-(x)$  for $H_-$.

From (\ref{171}) we obtain the following 
relation between $W(x)$ and $W_1(x)$
\begin{equation} \label{19}
W^2(x)+W'(x)=W_1^2(x)-W'_1(x) +2 \epsilon .
\end{equation}

Previously the same equation was used in the case of the so-called
shape invariant potentials to obtain the exact solutions of
Schr\"{o}dinger equation \cite{22} (see also reviews \cite{17,18}). 
We consider a more
general case and do not restrict ourselves to the 
shape invariant potentials.
Note, that (\ref{19}) is the Riccati equation 
which can not generally be solved exactly
with respect to $W(x)$ for a given $W_1(x)$ and 
vice versa.

The basic idea of this paper consists of finding
such a pair of $W(x)$ and
$W_1(x)$ that satisfies equation (\ref{19}).
It has been recently suggested by us in \cite{25}.
For this purpose
let us rewrite equation (\ref{19}) in the following form
\begin{equation} \label{20}
W'_+(x)=W_-(x)W_+(x) +2\epsilon,
\end{equation}
where
\begin{eqnarray} \label{21}
W_+(x)=W_1(x) + W(x),\\ \nonumber
W_-(x)=W_1(x) - W(x).
\end{eqnarray}
This new equation (\ref{20}) can be easily solved
with respect to
$W_-(x)$ for a given arbitrary function $W_+(x)$ 
or with respect to $W_+(x)$ for a given arbitrary function $W_-(x)$. 
Then from (\ref{21}) we obtain superpotentials $W(x)$ and $W_1(x)$ which
satisfy equation (\ref{19}).

\subsection{Solution with respect to $W_-(x)$}

In this subsection we construct the QES potentials using the solution 
of equation (\ref{20}) with respect to $W_-(x)$
\begin{equation} \label{22}
W_-(x)=(W'_+(x) - 2\epsilon )/W_+(x),
\end{equation}
where $W_+(x)$ is some function of $x$. 
Note, that the superpotentials 
$W(x)$ and $W_1(x)$ 
must satisfy condition (\ref{10}).
Then as one may see from (\ref{21}) $W_+(x)$ must satisfy the same 
condition (\ref{10})
as $W(x)$ and $W_1(x)$ do.

Let us consider continuous functions
$W_+(x)$. Because $W_+(x)$ satisfies 
condition (\ref{10}) the function $W_+(x)$ must pass through zeros. 
Then
as we see from (\ref{22}) $W_-(x)$, and thus $W(x)$,  
$W_1(x)$ have poles. 
In order to construct the superpotential free of singularities
suppose that $W_+(x)$ has only one zero at $x=x_0$
with the following behaviour in the vicinity of $x_0$
$
W_+(x)=W'_+(x_0)(x-x_0).
$
In this case the pole of $W_-(x)$ at $x=x_0$ can
be cancelled by choosing
\begin{equation} \label{25}
\epsilon = W'_+(x_0)/2.
\end{equation} 
Then the superpotentials free of singularities are
\begin{eqnarray}  \label{26}
W(x)={1\over 2}\left(W_+(x) - (W'_+(x)-W'_+(x_0))/W_+(x) \right), 
\\ \nonumber
W_1(x)={1\over 2}\left(W_+(x) + (W'_+(x)-W'_+(x_0))/W_+(x) \right). 
\end{eqnarray}

Substituting the obtained result for $W(x)$ into (\ref{221}) we obtain
QES potential $V_-(x)$ with explicitly known wave function of ground state
(\ref{9}) and wave function of the first excited state. 
The latter can be calculated using (\ref{17}) and (\ref{12})
\begin{equation} \label{27}
\psi_1^-(x)=C^-_1\ W_+(x) \exp\left(-\int W_1(x) dx\right).
\end{equation}  
It is indeed the wave function of first excited state 
because $W_+(x)$ has one zero.

We may choose various functions $W_+(x)$ and obtain as a result
various QES potentials. 
The functions $W_+(x)$ must be such
that $\psi_0^-(x)$ and $\psi_1^-(x)$ are square integrable. 
If the eigenfunctions $\psi_0^-(x)$ and $\psi_1^-(x)$ belong 
to the Hilbert space of square integrable functions in
which the Hamiltonian is Hermitian then these functions must be orthogonal
\begin{equation} \label{28}
<\psi^-_0|\psi^-_1>=-C_0^- C_1^-\left[\left.\exp\left(
-\int dx W_+(x)\right)\right]
\ \right|_{-\infty}^{\infty}=0.
\end{equation}
The wave functions must also satisfy appropriate
boundary conditions.

To conclude this subsection let us consider explicit examples.
Choosing  
$
W_+(x)=A\left({\rm sinh}(\alpha x)-{\rm sinh}(\alpha x_0) \right)
$
we obtain the well known QES potential derived in \cite{8,9} by
the method elaborated in the quantum theory of spin systems.
Note that the case $x_0=0$ corresponds to Razavy potential \cite{3}.
Following the proposed SUSY method this example
is considered in details in 
our earlier paper \cite{25}.

Consider the function $W_+(x)$ in the polynomial form 
\begin{equation}
W_+(x)=ax + bx^3, \ \ a>0,\ b>0
\end{equation}
which gives a new QES potential
\begin{equation}
V_-(x)={1\over 8}(a^2-12b)x^2+{ab\over 4}x^4+{b^2\over 8}x^6
+{3ab\over 8(a+bx^2)^2}+{3b\over 8(a+bx^2)}-{a\over 4}.
\end{equation}
The energy levels of ground and first excited states are
$E_0^-=0$, $E_1^-=a/2$. Note, that two energy levels of this potential
do not depend on the parameter $b$. The wave functions of those
states read
\begin{eqnarray}
\psi _0^-(x)=C_0^- (a+bx^2)^{3/4}e^{-x^2(2a+bx^2)/8}, \\ 
\psi _1^-(x)=C_1^- x(a+bx^2)^{1/4}e^{-x^2(2a+bx^2)/8}.
\end{eqnarray}
It is worth to stress that the case $b=0$ correspond to linear 
harmonic oscillator.

\subsection{Solution with respect to $W_+(x)$}

Equation (\ref{20}) is the first order
differential equation with respect to $W_+(x)$. 
A general solution can be written in the 
following form
\begin{equation}\label{29}
W_+(x)=\exp\left(
\int dx W_-(x)\right)\left[2\epsilon\int dx\exp\left(
-\int dx W_-(x)\right)+ \lambda \right],
\end{equation}
here $\lambda$ is the constant of integration.

In order to simplify solution (\ref{29}) let us choose
$W_-(x)$ to be of the form
\begin{equation}\label{30}
W_-(x)=-\phi''(x)/\phi'(x),
\end{equation}
and suppose that $\phi'(x)>0$.
Then
\begin{equation}\label{31}
W_+(x)=(2\epsilon\phi(x)+\lambda)/\phi'(x).
\end{equation}\
Note that the constant $\lambda$ can be included into the function 
$\phi(x)$ and thus for $W_+(x)$ we obtain
\begin{equation} \label{32}
W_+(x)=2\epsilon\phi(x)/\phi'(x).
\end{equation}
Finally for superpotentials $W(x)$ and $W_1(x)$ we have
\begin{eqnarray}\label{33}
W(x)=({1\over2}\phi''(x)+\epsilon\phi(x))/\phi'(x), \\ 
W_1(x)=({1\over2}\phi''(x)-\epsilon\phi(x))/\phi'(x). \label{34}
\end{eqnarray} 

Using this result for wave functions of the ground state
with the energy $E^-_0=0$ and excited state with $E^-_1=\epsilon$
we obtain
\begin{eqnarray} \label{35}
\psi^-_0(x)=C^-_0(\phi'(x))^{-1/2}\exp\left(
-\epsilon\int dx \phi(x)/\phi'(x)\right), \ \ E^-_0=0,\\ \nonumber
\psi^-_1(x)=C^-_1\phi(x)(\phi'(x))^{-1/2}\exp\left(
-\epsilon\int dx \phi(x)/\phi'(x)\right), \ \ E^-_1=\epsilon,
\end{eqnarray}
where function $\phi (x)$ must be such that these wave
functions are square integrable.
The condition of orthogonality in this case can by written similarly
to (\ref{28})
\begin{equation} \label{36}
<\psi^-_0|\psi^-_1>=-C^-_0C^-_1\left[\left.\exp\left(
-\epsilon\int dx \phi(x)/\phi'(x)\right)\right]
\ \right|_{-\infty}^{\infty}=0.
\end{equation}

QES potential $V_-(x)$ is given by (\ref{221}) 
with superpotential (\ref{33}).
Choosing different $\phi(x)$ we obtain different QES potentials
with explicitly known two eigenstates. We shall consider a nonsingular
monotonic function $\phi(x)$ with one node. Then $\psi^-_1(x)$
also has one node and thus corresponds to the first excited state.

In conclusion of this subsection let us consider an explicit example. 
Let us put
\begin{equation} \label{37}
\phi(x)=ax+bx^3/3,\ \ a, b>0.
\end{equation}
Note, that the case $b=0$ corresponds to
a linear harmonic oscillator.  
The function (\ref{37}) generates the following superpotentials
\begin{eqnarray}
W(x)=\epsilon x/ 3 +(b+2a\epsilon /3){x\over a+bx^2}, \label{371} \\
W_1(x)=\epsilon x/ 3 +(-b+2a\epsilon /3){x\over a+bx^2} \label{372}
\end{eqnarray}
which as we see satisfy condition (\ref{10}).

Substituting $W(x)$ into (\ref{221}) we obtain the following
QES potential $V_-(x)$ and its SUSY partner $V_+(x)$

\begin{eqnarray} \label{38}
V_-(x)={A_-\over2}x^2+{B_-\over a+bx^2}+{D_-\over (a+bx^2)^2}+R_-, \\
\label{39}
V_+(x)={A_+\over2}x^2+{D_+\over (a+bx^2)^2}+R_+,
\end{eqnarray}
where
\begin{eqnarray*}
A_-=A_+=\epsilon^2/9, \ \
B_-=b+{2\over3}a\epsilon, \ \
R_-=R_+={\epsilon\over18b}(3b+4a\epsilon), \\
D_-=-{1\over18b}(27ab^2+24a^2b\epsilon+4a^3\epsilon^2), \ \
D_+={1\over 18b}(9ab^2-4a^3\epsilon^2).
\end{eqnarray*}
Using (\ref{35}) we obtain the wave functions of the ground 
and first excited states 
\begin{eqnarray}
\psi^-_0(x)=C^-_0 \ (a+bx^2)^{-1/2-a\epsilon/3b}\ 
\exp(-\epsilon x^2/6)\label{40},\\
\psi^-_1(x)=C^-_1 \ (ax+bx^3/3)(a+bx^2)^{-1/2-a\epsilon/3b}\ 
\exp(-\epsilon x^2/6).\label{41}
\end{eqnarray}

The same result for QES potential (\ref{38}) and
wave functions (\ref{40}), (\ref{41}) can be obtained using the method
described in section 3.1 and taking the function $W_+(x)$
to be of the form (\ref{32}).

It is interesting to stress that in the special case 
$\epsilon=3b/2a$ the QES potential $V_-(x)$ reads
\begin{equation}
V_-(x)={b^2\over 8a^2}x^2+{2b\over a+bx^2}-{4ab\over (a+bx^2)^2}+
{3b\over 4a}
\end{equation}
and can be solved exactly.
To see this note that in this special case 
the superpotential $W_1(x)=\epsilon x/3$ corresponds to
superpotential of a linear harmonic oscillator.
Then $V^{(1)}_-(x)$ and, as a result of (\ref{171}),
$V_+(x)$ are the potential energies of the linear harmonic oscillator
\begin{equation}
V_+(x)={b^2\over 8a^2}x^2+{5b\over 4a}.
\end{equation}
The fact that $V_+(x)$ corresponds to the linear harmonic oscillator
follows also directly from (\ref{39}) because
the coefficient $D_+$ in the considered case is equal to zero.
Therefore in this case 
$H_+$ is the Hamiltonian of the linear harmonic oscillator and we know
all its eigenfunctions in explicit form.
Using SUSY transformations (\ref{11}), (\ref{12}) we can easily calculate
the energy levels and the wave functions of all the excited states of
$H_-$. 
Note that in this special case $V_-(x)$ can be treated as CES potential
and it corresponds to the one studied in \cite{19,20,23,24}. 

As far as we know the potential in general form (\ref{38}) 
has not been previously discussed 
in the literature. This potential is interesting from that 
point of view that in the case of
$\epsilon=3b/2a$ this potential is the CES one for which the whole
energy spectrum and the corresponding eigenfunctions can be calculated in
the explicit form.



\begin{thebibliography}{25}
\bibitem{1} V.~Singh, S.~N.~Biswas, K.~Dutta, Phys. Rev. D {\bf 18} 
            (1978) 1901.
\bibitem{2} G.~P.~Flessas, Phys. Lett. A {\bf 72} (1979) 289.
\bibitem{3} M.~Razavy, Am. J. Phys. {\bf 48} (1980) 285; Phys. Lett A 
{\bf 82} (1981) 7.
\bibitem{4} A.~Khare, Phys. Lett. A {\bf 83} (1981) 237.
\bibitem{5} A.~V.~Turbiner, A.~G.~Ushveridze, Phys. Lett. A {\bf 126} 
             (1987) 181.
\bibitem{6} A.~V.~Turbiner, Commun. Math. Phys. {\bf 118} (1988) 467.
\bibitem{7} M.~A.~Shifman, Int. Jour. Mod. Phys. A {\bf 4} (1989) 2897.

\bibitem{8} O.~B.~Zaslavskii, V.~V.~Ul'yanov, V.~M.~Tsukernik,
             Fiz. Nizk. Temp. {\bf 9} (1983) 511.
\bibitem{9} O.~B.~Zaslavsky, V.~V.~Ulyanov, Zh. Eksp. Teor. Fiz.
             {\bf 87} (1984) 1724.
\bibitem{10} D.~P.~Jatkar, C.~Nagaraja Kumar, A.~Khare, Phys. Lett. A 
             {\bf 142} (1989) 200.
\bibitem{11} P.~Roy, Y.~P.~Varshni, Mod. Phys. Lett. A {\bf 6}
             (1991) 1257.
\bibitem{12} A.~Gangopadhyaya, A.~Khare, U.~P.~Sukhatme,
             preprint hep-th/9508022 (1995).
\bibitem{13} V.~V.~Ulyanov, O.~B.~Zaslavskii, J.~V.~Vasilevskaya,
            Fiz. Nizk. Temp. {\bf 23} (1997) 110. 
\bibitem{14} A.~G.~Ushveridze, Quasi-exactly solvable models in quantum 
            mechanics,
            Institute of Physics Publishing, Bristol (1994).
\bibitem{15} A.~Krajewska, A.~Ushveridze and Z.~Walczak,
             Mod. Phys. Lett. {\bf A 12} (1997) 1225.
\bibitem{16} A.~Khare, B.~P.~Mandal, preprint quant-ph/9711001 (1997).
\bibitem{17} F.~Cooper, A.~Khare, U.~Sukhatme, Phys. Rep. {\bf 251} 
             (1995) 267.
\bibitem{18} G.~Junker, Supersymmetric methods in quantum and statistical
             physics (Springer, Berlin, 1996). 
\bibitem{19}  G. Junker, P. Roy, Phys. Lett. A {\bf 232} (1997) 155.
\bibitem{20} G. Junker, P. Roy, preprint quant-ph/9803024 (1998).
\bibitem{21} A. de Souza Dutra, Phys. Rev. A {\bf 47} (1993) R2435.
\bibitem{22} L.~E.~Gendenshteyn, Pisma Zh. Eksp. Teor.Fiz. {\bf 38}
             (1983) 299.
\bibitem{23} V.~G.~Bagrov, B.~F.~Samsonov, Teor. Mat. Fiz. {\bf 104}
             (1995) 356.
\bibitem{24} V.~G.~Bagrov, B.~F.~Samsonov, J. Phys. A {\bf 29}
             (1996) 1011.               
\bibitem{25} V.~M.~Tkachuk, preprint quant-ph/9801021 (1998). 
\end{thebibliography}
\end{document}